\documentstyle{article}[12pt]

\def\be{\begin{equation}}
\def\ee{\end{equation}}
\def\ben{\begin{displaymath}}
\def\een{\end{displaymath}}
\def\ba{\begin{array}{c}}
\def\ea{\end{array}}

\begin{document}


\vspace*{1cm}

\begin{center}
{\Large\bf
Quasi-exactly solvable quartic potentials

 with centrifugal
and Coulombic terms }
\end{center}


\begin{center}

Miloslav Znojil


OTF, \'{U}stav jadern\'e fyziky AV \v{C}R, 250 68 \v{R}e\v{z},
Czech Republic

homepage: http://gemma.ujf.cas.cz/\~\,znojil

e-mail: znojil@ujf.cas.cz

\end{center}

\vspace{5mm}



\section*{Abstract}

$D-$dimensional central and complex potentials of a Coulomb plus
quartic-polynomial form are considered in a ${\cal
PT}-$symmetrized radial Schr\"{o}dinger equation. Arbitrarily
large finite multiplets of bound states are shown obtainable in an
elementary form.  Relations between their energies and couplings
are determined by a finite-dimensional secular equation. The
Bender's and Boettcher's one-dimensional quasi-exact oscillators
re-emerge here as the simplest chargeless solutions.



\newpage

\noindent
\section{Introduction}

Recently, Bender and Boettcher \cite{BB} studied the
three-parametric family of Hamiltonians
 \be
 H^{(BB)} = -\partial^2_x - x^4 + 2\,i\,a\,x^3 + c\,x^2
 +i\,(a^3-a\,c -2\,J)\,x,
 \ \ \ \ \ \ a, c \in
 I\!\!R, \ \ \ \ \ J = 1, 2, \ldots\
 \label{Ham}
  \ee
defined on a certain complex curve $x=x(t) \in l\!\!\!C$, $t \in
(-\infty,\infty)$. Their main result was an explicit construction
of a $J-$plet of bound states in an elementary and closed form
 \ben
 \psi_j(x) =
  e^{-i\,x^3/3-a\,x^2/2-i\,b\,x}
P_{j,J-1}(x),\ \ \ \ \ \ \ \ \ b = (a^2-c)/2 , \ \ \ \ \ \ \ j =
1, 2, \ldots, J
 \ .
 \een
Here, $P_{j,J-1}(x)$ denotes certain polynomials of the $(J-1)$-st
degree. The normalizability of the multiplet is manifestly
guaranteed via pre-selection of the asymptotics of $x(t)$,
 \ben
 x(t) \sim
 \left (
  \cos \varphi \pm i\,\sin \varphi
 \right )
 t, \ \ \ \ \ \ \ t \to  \mp \infty\
 , \ \ \ \ \ \ \varphi \in (0,\pi/3).
 \een
It is easy to check that we may even stay on the straight line
$x(t) = t - i\,\varepsilon$, provided only that we choose it in
such a way that ${\rm Re}(2\,\varepsilon+a)>0$.

An older paper by Buslaev and Grecchi \cite{BG} may be recalled
for the rigorous proof that the Hamiltonian (\ref{Ham}) has the
real and discrete spectrum. The paper shows that after a constant
shift $\delta \sim b^2+J\,a/2$ the spectrum will coincide with the
energy levels of the $D-$dimensional anharmonic oscillator
$H^{(AHO)} = -\triangle + \frac{1}{4} \vec{r}^2 + \frac{1}{4}
\left [ \vec{r}^2\right ]^2$ in its $\ell-$th partial-wave mode
such that $2J = D+2\ell-2$. This assignment (or ``BG
transformation") combines a change of variables with Fourier
transformation. Its application to $H^{(BB)}$ at an integer $J$
specifies either our choice of the AHO partial wave $\ell = 0, 1,
\ldots$ or of the (even) dimension $D$. In the opposite direction
it throws new light on the non-Hermitian Hamiltonians $H^{(BB)}$
which exhibit the puzzling \cite{Bender} ${\cal PT}-$symmetry
 \ben
 {\cal PT} H^{(BB)}
= H^{(BB)}
 {\cal PT} ,
 \ \ \ \ \ {\cal P} \psi(x) = \psi(-x),
 \ \ \ \ \ {\cal T} \psi(x) = \psi^*(x).
 \een
The $J$-plets of their exact states become transformed in the
unusual anharmonic oscillator states obtainable in terms of
certain elementary Fourier-type integrals (cf. also ref.
\cite{Watt} in this context). This adds a new and strong reason
why the models of the type (\ref{Ham}) deserve a thorough
attention.

We may only feel dissatisfied by a certain incompleteness of the
whole picture: Why the BG transformation prefers the even
dimensions? What could be done at the odd $D= 1, 3, \ldots$? The
free variability of the integer $D$ would be highly welcome also
in some phenomenological $D \gg 1$ models in nuclear physics
\cite{Sotona}, quantum chemistry \cite{Groebner} and atomic
physics \cite{expan}.

The second reason why the class of models (\ref{Ham}) looks so
inspiring is related to the second version of the BG
transformation, presented in paper \cite{BG} as ``the second main
result". It starts from a less usual, {volcano-shaped} model
$V(\vec{r}) \sim \omega^2\vec{r}^2 - \left [ \vec{r}^2 \right
]^2$. The current partial-wave decomposition of its
$D-$dimensional wave functions
 \ben
 \Psi(\vec{r}) = \sum_{\ell=0}^\infty \ r^{(1-D)/2}\,\psi(r) \,
 \times angular \ part\ , \ \ \ \ \ \ r=|\vec{r}| \in (0, \infty)
 \een
reduces its Schr\"{o}dinger equation to the ordinary equation on
the half-axis. Using a  complex shift of the coordinates again,
Buslaev and Grecchi arrive at a new, properly ${\cal
PT}-$symmetrized two-parametric operator
 \be
 H^{(BG)} = -\partial^2_t +\frac{L(L+1)}{x^2}+
 \omega^2 x^2- x^4 \ , \ \ \ \ \ \ x= x(t) = t-i\,\varepsilon,
  \ \ \  L= \ell +(D-3)/2,
  \label{Hamba}
  \ee
extended to the whole real line $t \in I\!\!R$ and acting in the
current Hilbert space $L^2(I\!\!R)$. They prove its isospectrality
with a certain one-dimensional (and Hermitian) double-well
oscillator. In the resulting quadruple scheme
 \ben
  \ba
  \\
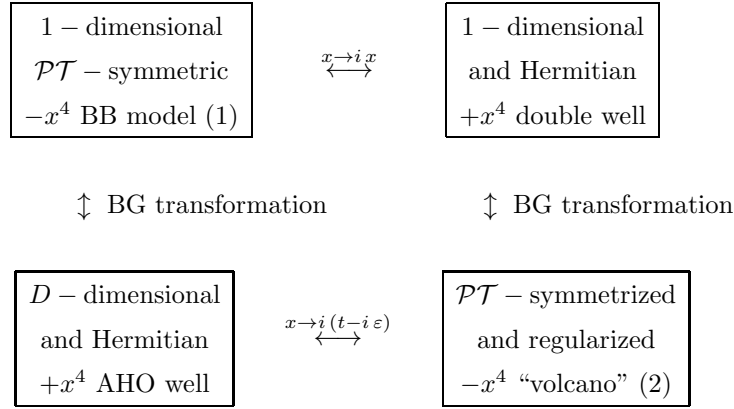
 \begin{array}{|c|}
 \hline
 1-{\rm dimensional}\\ {\cal PT}-{\rm symmetric}\\
 -x^4\  {\rm BB\ model}\ (\ref{Ham})\\
 \hline
 \ea \ \ \ \ \ \ \
 \stackrel{x \to i\,x}{ \longleftrightarrow }
  \ \ \ \ \ \ \
 \begin{array}{|c|}
 \hline
 1-{\rm dimensional}\\ {\rm and \ Hermitian}\\
 +x^4\  {\rm double\ well}\\
 \hline
 \ea
\\
\\ \ \ \ \ \ \ \ \ \ \ \ \ \ \ \ \  \ \
  \updownarrow \ {\rm BG\ transformation}\  \  \ \ \ \ \ \
 \ \ \ \
\ \ \ \ \  \updownarrow \ {\rm  BG\ transformation}\  \ \\
\\
 \ \ \ \
 \begin{array}{|c|}
 \hline
 D-{\rm dimensional}\\ {\rm and \ Hermitian}\\
 +x^4\  {\rm  AHO\ well}\\
 \hline
 \ea \ \ \ \ \
 \stackrel{x \to i\,(t-i\,\varepsilon)}{ \longleftrightarrow }
 \ \ \ \ \
  \begin{array}{|c|}
 \hline
 {\cal PT}-{\rm symmetrized}\\
 {\rm and \ regularized}\\
 -x^4\  {\rm ``volcano"}\ (\ref{Hamba})\\
 \hline
\ea
\\
\\
\ea
 \een
the two columns are related by the elementary changes of
variables. The upper-left-corner item is characterized by its
partial solvability. In what follows we intend to show that the
lower-right-corner-like {\em singular} Hamiltonians may equally
well exhibit the same (or at least very similar) multiplet or
``quasi-exact" \cite{Ushveridze} solvability.

\section{${\cal PT}-$regularized singular oscillators}

The problem we shall solve here is the Schr\"{o}dinger equation
 \be
\left[-\,\frac{d^2}{dx^2} + \frac{L(L+1)}{x^2}
 +i\,\frac{d}{x}+i\,c\,x
 +b\,x^2  + i\, a\,x^3 - x^4
 \right]\, \psi(x) =
E \psi(x)
 \label{SE}
 \ee
defined on curves $x(t)\neq 0$ to be specified later. It contains
the bound state problems (\ref{Ham}) and (\ref{Hamba}) after an
appropriate choice of its parameters. As long as certain
characteristic features of the similar ${\cal PT}-$symmetric
systems will be of an immediate relevance here, let us first
review some of them briefly.

\subsection{${\cal PT}-$symmetry}

Many complex models with ${\cal PT}$ symmetry (and, let us assume,
real spectra, Im $E=0$ \cite{Benderlin}) are defined on real line
\cite{Berlin}. Their potentials comprise a spatially symmetric
real well plus its purely imaginary antisymmetric complement. An
interpretation of these models parallels the usual real and
symmetric bound state problems in one dimension since one may try
to switch off the imaginary force smoothly. Typically, the
spectrum splits in the convergent and divergent parts in this
limit. Recalling eq. (\ref{Hamba}) for illustration and
definiteness, we identify the convergent part with the standard
physical solutions $\psi_{(qo)}(r) \sim r^{\ell+(D-1)/2}$ of a
``quasi-odd" or ``regular" type. The divergent part represents the
``quasi-even" components $\psi_{(qe)}(r) \sim r^{(3-D)/2-\ell}$.
For a further explicit illustration one may recall the harmonic
oscillator \cite{PTHO} or a few other solvable examples
\cite{Eckart}.

The less trivial set of the models which ``weaken" their
Hermiticity to the mere ${\cal PT}$ symmetry may be defined off
the real line \cite{Cannata}. In the usual mathematical notation
one speaks about sectors where the $L^2(I\!\!R)$ boundary
conditions are imposed. Thus, returning to our examples
(\ref{Ham}) -- (\ref{SE}) we define sectors $S_k=\{x \in l\!\!\!C;
x \neq 0,| \arg (x) -\frac{1}{6}(2k-1)\pi | < \frac{1}{6} \pi \}$
and choose $S_4$ and $S_6$ in accord with the recommendation of
ref. \cite{BB}, or $S_1$ and $S_3$ as accepted in ref. \cite{BG}
(cf. Figure 1).

The latter ambiguity of the curve $x(t)$ may be characterized by a
``signature" $\sigma = \pm 1$ which easily distinguishes between
the two ${\cal T}-$conjugate curves $ x(t)=x^{(\sigma)}(t)$ with
the asymptotics bounded by the $\varphi \to \pi/3$ Stokes lines,
 \be
 x^{(\sigma)}(t) \sim
 \left (
  \cos \varphi \pm i\,\sigma\,\sin \varphi
 \right )
 t, \ \ \ \ \ \ \ t \to  \mp \infty\
 , \ \ \ \ \ \ \varphi \in (0,\pi/3).
 \ee
Even in the admissible straight-line or $\varphi \to 0$ extreme
$x^{(\sigma)}(t) = t - i\,\sigma\,\varepsilon$, the possible
branch-point singularity in the origin must be avoided properly,
e.g., in the trivial ${\cal T}-$conjugation manner which leaves
the (real) spectrum unchanged. For this reason the condition which
determines $\varepsilon$ is now more restrictive and reads
$2\varepsilon>|a|$.

We may summarize that our curves of integration
$x(t)=x^{(\sigma)}(t)$ may be straight lines or hyperbolic shapes
open downwards ($\sigma = +1$, \cite{BB}) or upwards ($\sigma
=-1$, \cite{BG}). For the fully regular forces their
non-asymptotic form may be deformed almost arbitrarily. For all
the other models with singularities these curves must avoid all
the cuts. In particular, eqs. (\ref{Hamba}) and (\ref{SE}) admit
the presence of an essential singularity in the origin. After we
cut the plane upwards (for $\sigma = +1$) or downwards (for
$\sigma = -1$), both our curves $x=x^{(\sigma)}(t)$ may be kept
smooth and, say, symmetric with respect to the ${\cal T}$
reflections.

As long as we are relaxing the usual Hermiticity, $H^{(BB)} \neq
[H^{(BB)}]^+$, both the two independent wave function solutions
remain equally admissible in the vicinity of the origin. This is
the main formal reason why the Bender's and Boettcher's complex
oscillator remained solvable: One only has to combine the
quasi-even behaviour $\psi_{(qe)}(r) \sim {\cal O}(r^0)$ with its
quasi-odd parallel $\psi_{(qo)}(r) \sim {\cal O}(r^1)$ in the
single ansatz. Here we intend to proceed in the same manner.

\subsection{Recurrences}

Schr\"{o}dinger equation (\ref{SE}) with the integer or
half-integer values of $L= \ell +(D-3)/2=K/2$ remains perfectly
regular along both our non-self-intersecting integration paths.
After a change of variables $x^{(\sigma)}(t)= -i\,y(t)$ we get the
new differential equation for $\psi(x) = \chi(i\,x)$ which
contains, formally, no imaginary units,
 \be
\left[\,\frac{d^2}{dy^2} - \frac{L(L+1)}{y^2}
 -\frac{d}{y}+c\,y
 -b\,y^2  -  a\,y^3 - y^4
 \right]\, \chi(y) =
E \chi(y).
 \label{SEre}
 \ee
We shall search for its solutions by a power-series ansatz,
assuming that such a series terminates. Demanding also the
manifest normalizability in a way depending on the signature
$\sigma = \pm 1$ we arrive at a virtually unique formula
 \be
\chi(y)=\exp \left( \sigma\, \frac{1}{3} \,y^3 +
 \frac{1}{2} T \,y^2 +S\,y \right)\
\sum^{N}_{n=0} h_n\,y^{n-L}.
 \label{anhari}
 \ee
It combines both the respective $y^{-L}$ and $y^{L+1}$ quasi-even
and quasi-odd components, and equation (\ref{SEre}) determines the
asymptotically correct values of our auxiliary parameters as well
as the unique termination-compatible coupling $c$,
 \ben
 T=a/2\sigma,\ \ \ \ \ \
 S=(b-T^2)/2\sigma, \ \ \ \ \ \
 c \equiv c(N) = -2TS-\sigma\cdot(2N-2L+2).
 \een
The resulting ansatz (\ref{anhari}) represents the desired
normalizable solutions if and only if its coefficients $h_n$ are
compatible with the recurrences
 \be
 h_{n+1} A_{n} + h_{n}B_{n}+
h_{n-1} C_{n} + h_{n-2}D_{n}
 = 0 , \ \ \ \ \ \ \
  n =  0, 1, \ldots, N+1.
  \label{4.8}
 \ee
Their coefficients
 \ben
  A_n= (n+1)(n-2L),
  \ \ \ \
 B_n= S(2n-2L)-d,
  \ \ \ \
 C_n= S^2+T(2n-2L-1)-E,
  \ \ \ \
 D_n= 2\sigma(n-N-2)
 \een
are all elementary.

\section{Finite-dimensional Schr\"{o}dinger equation}

In the light of their strict postulated termination, our
recurrences (\ref{4.8}) represent just a finite set of $N+2$
linear algebraic equations for $N+1$ unknown coefficients $h_n$.
Such a set is, obviously, over-determined. Its $(N+2)\times
(N+1)-$dimensional non-square matrix has in effect two main
diagonals,
 \ben
 B_n= S(2n-K)-d,
  \ \ \ \ \ \ \ \ \ \ \ \ \ \ \
 C_n= S^2+T(2n-K-1)-E, \ \ \ \ \ \ K = 2L.
 \een
Both the quantities $d$ and $E$ can play a role of an eigenvalue
{\em simultaneously} \cite{Dubna}. In the other words, we may
interpret the whole linear set of equations as the two coupled
square-matrix problems indicated, say, by the single and double
line in the whole non-square system
 \be
 \left(
\begin{array}{ccccc}
\hline
 B_0&A_0&&&\\ \hline \hline
 C_1&B_1&A_1&&\\
   D_2&C_2&\ddots&\ddots&\\
    &\ddots&\ddots&B_{N-1}&A_{N-1}\\
 &&D_N&C_N&B_{N}\\
 \hline
 &
&&D_{N+1}&C_{N+1}\\
 \hline \hline
 \ea \right )
\left ( \ba h_0\\ h_1\\ h_2\\ \vdots \\ h_N
 \ea
\right ) =0\ .
 \label{err4.8}
 \ee
In both cases (i.e., omitting the last or first line,
respectively) we get a routine secular determinant with an
eigenvalue on its main diagonal. Both these two independent
secular or termination conditions have to be satisfied
simultaneously.

\subsection{Auxiliary constraint}

As long as the upper diagonal $A_n= (n+1)(n-K)$ vanishes at $n=K$,
there emerges an important asymmetry between our two eigenvalues
$d$ and $E$. Whenever one keeps just a few lowest partial waves,
we may say that the integer $K=2L$ remains ``very small", at least
in comparison with $N$ which can/should be ``large" or at least
``arbitrary" in principle. This is our present fundamental
observation. As a consequence of the related disappearance of
$A_K=0$ we shall achieve a thorough simplification of the
construction of bound states.

At the lowest possible $K=0$ we return immediately to the Bender's
and Boettcher's proposal. Their choice of $d=0$ in eq. (\ref{Ham})
(with $J=N+1$) gives us the highly welcome possibility of omitting
the whole first row from eq. (\ref{err4.8}).  The rest of this
equation is the usual square-matrix diagonalization. All the
exceptional quasi-exact eigenvalues $E_j$, $j = 1, 2, \ldots, N+1$
may be determined as roots of a polynomial of the $(N+1)-$st
degree \cite{BB}.

A transition to the nonzero integers $K$ is easy. In place of
using the trivial $d=0$ we have to satisfy the $(K+1)-$dimensional
sub-equation
 \be \det \left(
\begin{array}{ccccc}
B_0&A_0&&&\\
 C_1&B_1&A_1&&\\
   D_2&C_2&\ddots&\ddots&\\
    &\ddots&\ddots&B_{K-1}&A_{K-1}\\
 &&D_K&C_K&B_{K}
 \ea \right )
=0
 . \label{eq4.8}
 \ee
This can fix, say, the {eligible} electric charges $d$ as
functions of the other parameters. These functions can be used as
a starting point of a facilitated solution of our original problem
(\ref{err4.8}).

\subsection{Closed formulae for the energies}

In the simplest test of our new and general recipe let us return,
once more, to $K=0$ in eq. (\ref{eq4.8}) and derive
 \ben
 d = 0, \ \ \ \ \ \ K=0.
 \een
This reproduces the model $H^{(BB)}$ of ref. \cite{BB}. In the
next (and the first really innovative) $K=1$ step we get the
double root $ d=d_{(\pm)}(E) = \pm \sqrt{E}$. Each choice of the
sign specifies a different relation between the charge and the
energy. Both these signs remain admitted by the inverted, unique
recipe
 \ben
E=E_K(d)=d^2, \ \ \ \ \ \ K=1.
 \een
It defines the energy as a function of the charge. The numbering
of the separate elements of our new, $K=1$ multiplets of bound
states is in this way transferred directly to the admissible
charges $d=d_j$. Formally this means that the energy $E$ may be
eliminated from all our algebraic equations. The values of the
charge $d$ remain the only unknown quantities.

At the subsequent integer $K=2$ we get, with a bit of luck, a
highly compact energy formula again,
 $$ E=E_K(d)=
 \frac{d^2}{4} +
 2\,\frac
{S\,T+\sigma\,N}{d}, \ \ \ \ \ \ \ \ K = 2.
 $$
The further step gives the two rules or roots $E_3(d)=F$ of the
quadratic equation
 $$ 9\,{F}^{2}-10\,{d}^{2}F
+{d}^{4}+48\,dST+48\,d\,\sigma\,N -24\,d-72\,S-36\,{T}^{2 }-0, \ \
 \ \ K=3. $$
Hence, all the $K \geq 3$ cases require the so called Gr\"{o}bner
elimination which should (and easily can) be performed by a
computer. We only quote a similar calculation \cite{Groebner} and
omit the further details.

\section{Secular determinant and its roots}

In the usual quasi-exact manner \cite{Ushveridze} the multiplets
of the (real) values of $d_j$ have to follow from our termination
postulate at any positive integer $N$. Let us now assume that the
value of $K$ is fixed. After the insertion of its energy formula
$E = E_K(d)$ in the recurrences or equation (\ref{err4.8}) we are
left with the square-matrix secular equation of the dimension
$(N+1)\times(N+1)$,
 \ben \det \left(
\begin{array}{ccccc}
 C_1(d)&B_1(d)&A_1&&\\
   D_2&C_2(d)&\ddots&\ddots&\\
    &\ddots&\ddots&B_{N-1}(d)&A_{N-1}\\
 &&D_{N}&C_{N}(d)&B_{N}(d)\\
 &&&D_{N+1}&C_{N+1}(d)
 \ea \right )
=0 \ . \label{eq40}
 \een
This equation has to determine the unknown eigencharges. In its
analysis we may skip the already known Bender's and Boettcher's
$K=0$ case \cite{BB} and pay attention to the next few $K= 1, 2,
\ldots$.

\subsection{The first nontrivial option: $K=1$}

Insertion of the available formulae gives us the $K=1$ secular
determinants with the only $N-$ and $\sigma-$dependent elements
$D_j(N)=2\sigma(n-N-2)$,
 \be \det \left(
\begin{array}{cccccc}
 S^2-d^2&S-d&0&&&\\
   D_2(N)&S^2+2T-d^2&3S-d&3&&\\
    &D_3(N)&S^2+4T-d^2&\ddots&\ddots&\\
 && \ddots\ \ \ \ \ \ \ \ \ &
   \ddots \ \ \ \ \ \ \ \ &&\\
   &&\ \ \ \ \ \ \ \ D_{N+1}(N)
   &\multicolumn{3}{r}{\ \ \ \ \ S^2+2NT-d^2}
 \ea \right ) =0 \ . \label{equad40}
 \ee
At any $N$ this equation gives an ``exceptional" explicit root
$d=S$. Although it looks like an artifact of our present approach,
its further inspection reveals its acceptability. For example, at
$N=1$ this root gives a solution provided that we have
$ST=-\sigma/2$.

All the other roots of eq. (\ref{equad40}) remain manifestly
$N-$dependent. They also exhibit certain symmetries. For example,
the change of the signature $\sigma \to -\sigma$ may be
compensated by the simultaneous sign-change of $S \to -S$ and $d
\to -d$.

In a more constructive setting it makes sense to pick up a
specific $N$. Starting from the smallest $N=1$ we arrive at the
cubic equation
 \ben
 -d^3-S\,d^2+(S^2+2T)\,d+2\,\sigma+S\,(S^2+2T)=0, \ \ \ \
 \ \ \ \ N=1.
 \een
We may determine the $S-$ and $T-$dependence of its three roots
$d_1, \ d_2$ and $d_3$ via Cardano formulae. If needed, the
boundaries of the domain of their reality may be determined
numerically, in a complete parallel to the $K=0$ study \cite{BB}.
In order to prove that this domain is not empty at $K=1$, we may
recall a sample triplet, say, of $d_{1,2,3,}(S, T) =(
-5.303953910, -3.103253421, 5.407207331)$ at $(S,T) = (3,10)$.

The growth of the dimension $N+1$ makes the practical
determination of the roots $d$ of eq. (\ref{equad40}) more
complicated. This is well illustrated by its next two $K=\sigma=1$
explicit polynomial representants,
 $$
 \ba
{d}^{5}+S{d}^{4} -\left (2\,{S}^{2}+6\,T\right ){d}^{3} -\left
(6+2\,{ S}^{3}+6\,ST\right ){d}^{2}
 +\left
(6\,{S}^{2}T+4\,S+8\,{T}^{2}+{S}^{4} \right )d\\
+10\,{S}^{2}+16\,T+6\,{S}^{3}T+8\,S{T}^{2}+{S}^{5} =0, \ \ \ \ \
 \ \ N = 2
 \ea
 \label{sect}
 $$
 $$ \ba -{d}^{7}-S{d}^{6}+\left (12\,T+3\,{S}^{2}\right
){d}^{5}+\left (12+12 \,ST+3\,{S}^{3}\right ){d}^{4}
 -\left
(16\,S+44\,{T}^{2}+3\,{S}^{4}+24 \,{S}^{2}T\right ){d}^{3}\\
-\left (40\,{S}^{2}+88\,T+44\,S{T}^{2}+3\,{S
}^{5}+24\,{S}^{3}T\right ) {d}^{2}\\ +\left
(12+16\,{S}^{3}+{S}^{6}+44\,{S
}^{2}{T}^{2}+64\,ST+48\,{T}^{3}+12\,{S}^{4}T\right )d\\
+152\,{S}^{2}T+28 \,{S}^{4}+84\,S+144\,{T}^{2}+{S}^{7}
+44\,{S}^{3}{T}^{2}+48\,S{T}^{3}+ 12\,{S}^{5}T =0
 , \ \ \ \ \ \ \ N = 3.
 \ea
 $$
To this list with the respective 16 and 31 terms one could add the
next $N=4$ item containing 53 terms, etc. This would complement
the similar $K=0$ formulae displayed in detail in ref. \cite{BB}.

\subsection{Numerical illustration }

An overall insight in the structure of the spectra may be mediated
by their simplest $S=T=0$ sample. {\it A posteriori}, the main
merit of such a choice may be seen in a drastic formal reduction
of the underlying secular polynomials. One may factorize many of
them by the purely non-numerical means.

\subsubsection{$K=0$}

In the paper \cite{BB} the domain of existence of the full
quasi-exact $N-$plets with $K=0$ has been determined numerically.
This analysis excluded the point $S=T=0$ where, empirically, the
${\cal PT}$ symmetry becomes spontaneously broken.
Computationally, this means that some of the energies coalesce and
move in pairs off the real line. Hence, our understanding of the
$K=S=T=0$ spectra is not complete yet.

Direct computation at $N=0$, $N=1$ and $N=2$ gives just the
trivial $E=0$. In the next two cases we get the
$\sigma-$independent equations
 \ben
 \ba
 E^4-96\,E=0, \ \ \ \ \ \ \ \ N = 3\\
 -E^5+336\,E^2=0, \ \ \ \ \ \ \ \ N = 4
 \ea
 \een
with the single nonzero real root $E \sim 4.48$ at $N=3$ and
 $E\sim 6.95$ at $N = 4$. For the general $N$ the secular
 equation reads
 \ben
 \det
 \left (
 \begin{array}{cccccc}
 -E&0&1\cdot 2&&&\\
 2N\sigma&-E&0&2\cdot 3&&\\
 & \ddots &\ddots&\ddots&\ddots& \\
 &&6\sigma&-E&0&(N-1)N\\
 &&&4\sigma&-E&0\\
 &&&&2\sigma&-E
 \ea
 \right )=0.
 \een
Its $\sigma-$independence and polynomiality in $x=E^3 \neq 0$ can
be proved in an easy exercise. As a consequence, its roots remain
non-numerical up to $N = 13$. In the latter extreme we get the
real quadruplet of nonzero energies $E \sim 9.381,\ 17.768,\ 26.\
487$ and $ 35.535$ .

\subsubsection{$K=1$}

After one moves to the negative signature $\sigma = - 1$, only
certain signs change in the secular polynomials of sect.
\ref{sect}. At both $\sigma=\pm 1$ we may reduce the first few
cases to their pertaining $S=T=0$ forms. They remain non-numerical
and exactly solvable up to the dimension $N=6$. Their first few
samples are
 \ben
 -d^3+ 2 \sigma= 0 , \ \ \ \ \ \ \ N=1
 \een
\ben
 d^5- 6\,\sigma\,d^2= 0 , \ \ \ \ \ \ \ N=2
 \een
\ben
 -d^7+ 12\,\sigma\,d^4+12\,d = 0 , \ \ \ \ \ \ \ N=3
 \een
\ben
 d^9 - 20\,\sigma\,d^6-76\,d^3+ 512\,\sigma\, = 0 , \ \ \ \ \ \ \
 N=4.
 \een
In detail, the single real root $d \approx 1.26 \cdot \sigma$ is
obtained at $N=1$, and one double zero and one real nonzero root
$d \approx 1.71 \cdot \sigma$ appear at $N=2$. One simple zero and
one positive and one negative root ($d_1 \approx 2.35 \cdot
\sigma,\ d_2 \approx -0.975 \cdot
 \sigma$) follow
 at $N=3$ while, finally,
 three nonzero real roots
$\sigma \cdot d_{1,2,3} \approx (2.82,\, 1.55,\,-1.83)$ emerge
from our last displayed polynomial at $N=4$.

\subsubsection{$K=2$}

In a move to the higher $K>1$ one has to notice that the first few
choices of $N \leq K$ are too formal and do not make an explicit
use of the present ``advantage" $A_K=0$ at all. In this sense, the
$K=2$ illustration has to start from the first nontrivial $N=3$.
After we fix $\sigma=+1$ for brevity, we get the secular
polynomial of the deterring twelfth degree. Fortunately, in the
same spirit as observed above the abbreviation $x=d^3\neq 0$
reduces it again to a (still solvable) quartic equation,
 $$ {x}^{4}+331776-96\,{x}^{3}+384\,{x}^{2}+18432\,x=0,
  \ \ \ \ \ \sigma=1,
 \ \ \ \ \ K=2,\ \ \ \ N=3\ .
 $$
This equation possesses the two real and positive roots, namely,
$x_1=24$ and
 $$ x_2=24+16\,\sqrt [3]{9+\sqrt
{17}}+{\frac {64}{\sqrt [3]{9+\sqrt {17}}}}\ \approx 88.87294116.
 $$
These roots lead to the real and positive charges
 $d_{1,2}\approx (2.88, \, 4.46)$
 and to the $K=2$ energies $E_2(d_j)$ as prescribed above.
The parallel problem with $\sigma = -1$
 leads to the full quadruplet of the
negative real roots
 $$x_{1,2,3,4}\approx (-199.78, -72.00, -14.65, -1.57),
 \ \ \ \ \ \ \sigma = -1 $$
 and to the eigencharges with the same signs,
 $d_{1,2,3,4}\approx (-5.84, -4.16, -2.45, -1.16)$.

We may conclude that at a fixed $K$ (i.e., for a specific partial
wave $\ell$) we get in general a multiplet of states connected by
some broken lines in the two-dimensional charge-energy plane. Only
in the simplest $K=0$ special case this ``guide for eye" becomes
the Bender-Boettcher straight line $d=const=0$.

\section{Discussion}

We have shown that the $N-$plets of exact bound-state solutions of
our general quartic problem (\ref{SE}) can be constructed in
closed form at any integer degree $N$ and dimension $D$. In the
other words, under certain relationship between the couplings and
energies, arbitrarily large multiplets of solutions (with real
energies!) proved obtainable from a single and finite-dimensional
secular equation. The Bender's and Boettcher's quartic example
re-appears here as the simplest one-dimensional special case with
the unique (and, incidentally, vanishing) Coulombic charge $d$.

In many applications of practical interest the closed quasi-exact
solutions are of the similar Sturmian type. Their bound states are
numbered by one of the couplings but still share, mostly, the same
value of the energy. Many Hermitian models belong to this
category, be it one of the most popular non-polynomial examples
\cite{Whitehead} or one of the historically first models of the
quasi-exact type \cite{Hautot}. Only the most popular partially
solvable sextic model \cite{Singh} owes for its popularity to its
most standard variable-energy character.

We have presented the explicit construction of a few multiplets
which lie along the more general curves (e.g., parabolas) in the
energy-coupling space. In combination with their overall ${\cal
PT}$ symmetric quantum mechanical framework these multiplets offer
several new challenges, e.g., in their completely missing abstract
interpetation, say, from the modern Lie-algebraic point of view
\cite{Turbiner}.

In our present approach the recent unexpected construction of
Bender and Boettcher \cite{BB} finds one of its simplest and quite
natural explanations. We have found that its solvable status has
been mediated first of all by its complex, non-Hermitian
background. In a way which clarifies the whole paradox the
exceptional features of the Bender's and Boettchere's oscillator
were related to its special chargeless form.

In another approach to the same problem one may just speak about
the underlying system of recurrences. In this language, it
remained unnoticed in the current literature on anharmonic
oscillators that the coefficient $A_{K}$ can vanish at the integer
$K = 2L = 2\ell + D -3$. Still, in one dimension, this was just
the core of feasibility of the Bender's and Boettcher's surprising
construction. This immediately implies that the Bender's and
Boettcher's one-dimensional potential is just a very special case
of the general class of quasi-exactly solvable quartic models.
Their description has been offered here.

It is probably worth re-emphasizing that the vanishing of the
coefficient $A_{K}=0$ does occur for the $s-$waves in three
dimensions as well as for $p-$waves in one dimension. This is our
re-interpretation of the known $K=0$ results. In the same sense,
the present new $K=1$ case covers the $s-$waves in four dimensions
and $p-$waves in two dimensions. Similarly, we have $K=2$ for
$s-$waves in five dimensions and for $p-$waves in three
dimensions, and we encounter the triple possibility of
$(\ell,D)=(0,6), \, (1,4)$ and $(2,2)$ at $K=3$. Etc. The first
problem formulated in our introductory section is satisfactorily
settled: In principle, the multiplet solvability occurs in all
dimensions.

Our second initial motivation concerned the possible tractability
of strong singularities of the centrifugal and Coulombic type. We
have preserved their certain constrained variability within the
Buslaev's and Grecchi's ${\cal PT}$ regularization scheme. In this
way we achieved a satisfactory balance in the picture given at the
end of section 1.

Further lessons from  our constructdion are not quite clear. In
the future, we intend to pay more attention to the underlying
Fourier-mediated $p \leftrightarrow x $ symmetries, trying to move
beyond their known harmonic-oscillator, BG-transformation
\cite{BG} or quasi-harmonic-oscillator \cite{SAO}
exemplifications. In such a context, the role of the
centrifugal-like singularities does not seem to have said its last
word yet.

\section*{Acknowledgements}

My thanks belong to Rajkumar Roychoudhury (ISI Calcutta) and to
Francesco Cannata (INFN Bologna) who insisted that the
Bender-Boettcher model deserves a deeper study. Partial support by
the GA AS CR grant Nr. A 1048004 is acknowledged.

\end{document}